\documentclass[final,epj]{svjour}
\usepackage{graphics}

\begin{document}

\title{Comparison among the local atomic order of amorphous TM-Ti alloys (TM=Co, Ni, Cu) 
produced by Mechanical Alloying studied by EXAFS}

\author{K. D. Machado \and J. C. de Lima \and C. E. M. Campos \and T. A. Grandi}

\titlerunning{Comparison among the local atomic order of amorphous TM-Ti alloys 
studied by EXAFS}

\authorrunning{K. D. Machado et al.}

\institute{Departamento de F\'{\i}sica, Universidade Federal de Santa Catarina, 88040-900 
Florian\'opolis, SC, Brazil}

\offprints{K. D. Machado}
\mail{kleber@fisica.ufsc.br}

\abstract{
We have investigated the local atomic structure of amorphous TM-Ti alloys (TM = Co, Ni, Cu) 
produced by Mechanical Alloying by means of EXAFS analyses on TM and Ti K-edges. Coordination numbers 
and interatomic distances for the four alloys where found and compared. EXAFS results obtained indicated 
a shortening in the unlike pairs TM-Ti as the difference between $d$ electrons of TM and Ti atoms 
increases, suggesting an increase in the chemical short range order (CSRO) from TM = Co to Cu.
}

\maketitle

\section{Introduction}

Mechanical alloying (MA) technique \cite{MA} is an efficient method for synthesizing crystalline 
\cite{SeZn,CarlosFeSe,CarlosCoSe,Joao1} and amorphous \cite{KleCoTi,KleNiTi,Joao2,Weeber,CarlosGaSe} 
materials, as well as stable and metastable solid solutions \cite{Froes,Yavari}. MA 
has also been used to produce materials with nanometer sized grains and alloys whose components have 
large differences in their melting temperatures and are thus difficult to produce using techniques 
based on melting. It is a dry milling process in which a metallic or non-metallic powder mixture 
is actively deformed in a controlled atmosphere under a highly energetic ball charge. The few 
thermodynamics restrictions on the alloy composition open up a wide range of possibilities for 
property combinations \cite{Poole}, even for immiscible elements \cite{Abbate}. 
The temperatures reached in MA 
are very low, and thus this low temperature process reduces reaction kinetics, allowing the 
production of poorly crystallized or amorphous materials.

We have used MA to produce three amorphous TM-Ti alloys:  
Co$_{57}$Ti$_{43}$ ({\em a}-Co$_{57}$Ti$_{43}$), Ni$_{60}$Ti$_{40}$ ({\em a}-Ni$_{60}$Ti$_{40}$) and 
Cu$_{64}$Ti$_{36}$ ({\em a}-Cu$_{64}$Ti$_{36}$) starting from the crystalline elemental powders. 
In Ref. \cite{KleCoTi} {\em a}-Co$_{57}$Ti$_{43}$ was studied by EXAFS and x-ray diffraction.
In Ref. \cite{KleNiTi} we found coordination numbers and interatomic distances 
for the first neighbors for {\em a}-Ni$_{60}$Ti$_{40}$ using EXAFS and RMC simulations 
\cite{RMC1,RMCA,rmcreview} of its total structure factor ${S}(K)$. Here, we studied 
{\em a}-Cu$_{64}$Ti$_{36}$ by EXAFS, and coordination numbers and interatomic distances were found and 
compared to those found for the other alloys. Due to its selectivity and high sensitivity to the chemical 
environment around a specific type of atom of an alloy, EXAFS is a technique (good reviews are found in 
Refs. \cite{Lee,Teo,Hayes,Prins}) very suitable to investigate the local atomic order of crystalline 
compounds and amorphous alloys. Anomalous wide angle x-ray scattering (AWAXS) is also a selective technique, 
but due to the small $K_{\rm max}$ that can be achieved on Ti K-edge ($\sim 4$ \AA$^{-1}$), little 
information could be obtained from an AWAXS experiment at the Ti K-edge. On the other hand, EXAFS 
measurements performed on this edge extended up to $\sim 16$ \AA$^{-1}$ in some alloys, allowing the 
determination of structural data with reasonable accuracy. Using this technique, we have determined 
coordination numbers and interatomic distances in the first coordination shell of {\em a}-Cu$_{64}$Ti$_{36}$. 
In addition, the chemical short range order (CSRO) in the three TM-Ti alloys were compared and it 
increases as TM goes from Co to Cu, in accordance with results reported by Hausleitner and Hafner using MD 
simulations \cite{Hausleitner}.

\section{Experimental Procedure}

Blended TM (TM = Co, Ni and Cu) and Ti crystalline elemental powders (Co:Vetec, 99.7\%, 
particle size $< 10$ $\mu$m; Ni: Merck, 99.5\%, particle size $< 10$ $\mu$m; Cu: Vetec 99.5\%, 
particle size $< 10$ $\mu$m; Ti: BDH, 99.5\%, particle size $< 10$ $\mu$m), with initial nominal 
compositions TM$_{60}$Ti$_{40}$, were sealed together with several steel balls, under an argon atmosphere, 
in a steel vial (more details can be seen at Refs. \cite{KleCoTi} and \cite{KleNiTi}). 
The ball-to-powder weight ratio was 5:1 for the four alloys. The vial was mounted in a Spex 8000 shaker 
mill and milled for 9 h. A forced ventilation system was used to keep the vial temperature close to room 
temperature. The composition of the as-milled powder was measured using the Energy Dispersive 
Spectroscopy (EDS) technique, giving the compositions Co$_{57}$Ti$_{43}$, 
Ni$_{60}$Ti$_{40}$ and Cu$_{64}$Ti$_{36}$, and impurity traces were not observed. EXAFS measurements were 
carried out on the D04B beam line of LNLS (Campinas, Brazil), using a channel cut monochromator (Si 111), 
two ionization chambers as detectors and a 1 mm entrance slit. This yielded a resolution of about 1.6 eV 
at the Ti K edge and 3.9 eV at the Co, Ni and Cu  K edges. 
All data were taken at room temperature in the transmission mode. The energy and average 
current of the storage ring were 1.37 GeV and 120 mA, respectively.

\section{Results and Discussion}

\begin{figure}[h]
\includegraphics{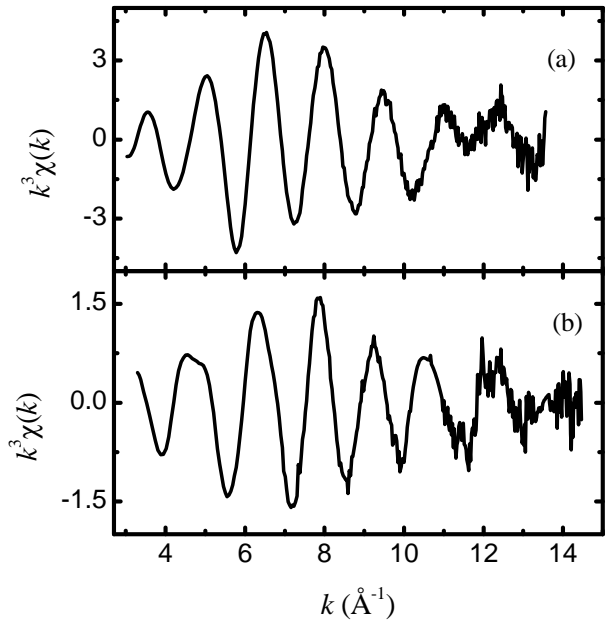}
\caption{\label{fig1}Weighted experimental EXAFS spectra: (a) at the Cu K edge and (b) at the 
Ti K edge.}
\end{figure}

\begin{figure}[h]
\includegraphics{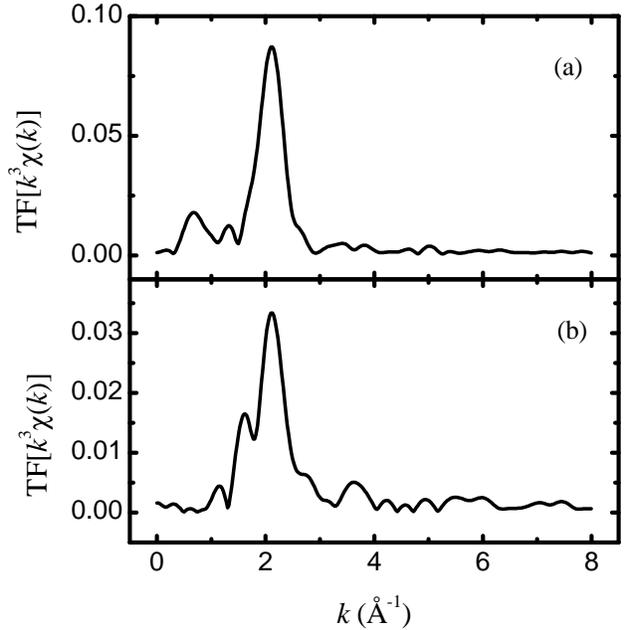}
\caption{\label{fig2}Fourier transformation of experimental EXAFS spectra: a) at the Cu K-edge and 
b) at the Ti K-edge.}
\end{figure}

The EXAFS oscillations $\chi(k)$ on Cu and Ti K edges of {\em a}-Cu$_{64}$Ti$_{36}$ are shown in 
Fig. \ref{fig1} weighted by $k^3$. After standard data reduction procedures using Winxas97 
software \cite{Ressler}, 
they were filtered by Fourier transforming $k^3 \chi (k)$ on both edges (Cu edge, 3.25 -- 13.6 \AA$^{-1}$ 
and Ti edge, 3.5 -- 14.5 \AA$^{-1}$) using a Hanning weighting function into $r$-space 
(Fig. \ref{fig2}) and transforming back the first coordination shells (1.30 -- 2.67 \AA\ for Co edge and 
1.85 -- 3.24 \AA\ for Ti edge). Filtered spectra were then fit by using Gaussian distributions to represent 
the homopolar and heteropolar bonds \cite{Stern}. We also used the third cumulant option of Winxas97 to 
investigate the presence of asymmetric shells. The amplitude and phase shifts relative to the homopolar 
and heteropolar bonds needed to fit them were obtained from ab initio calculations using the spherical waves 
method \cite{Rehr} and FEFF software.

\begin{figure}
\includegraphics{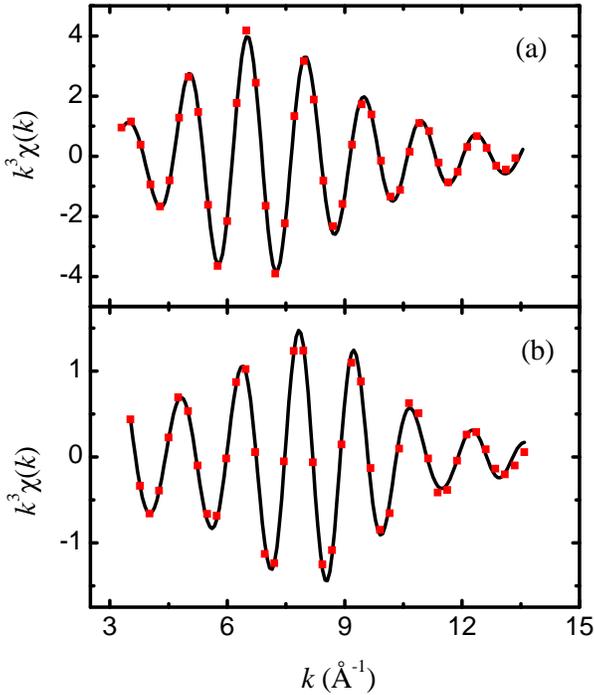}
\caption{\label{fig3}Fourier-filtered first shell (full line) and its simulation (squares) for 
{\em a}-Cu$_{64}$Ti$_{36}$ at the (a) Cu K edge, (b) Ti K edge.}
\end{figure}

Figure~\ref{fig3} shows the experimental and the fitting results for the Fourier-filtered first 
shells on Cu and Ti edges. Structural parameters extracted from the fits, including the errors 
in these values, are listed in table~\ref{tab1}. 
This table also shows data concerning crystalline Cu$_2$Ti 
({\em c}-Cu$_2$Ti, JCPDS card N$^{\rm o}$ 200371). It can be seen from Table \ref{tab1} that the number of 
Cu-Cu pairs in {\em a}-Cu$_{64}$Ti$_{36}$ is much higher than it is in {\em c}-Cu$_2$Ti, whereas the Cu-Cu 
average interatomic distance increases in the amorphous alloy. There is a shortening in the Cu-Ti 
distance in 
{\em a}-Cu$_{64}$Ti$_{36}$, which becomes shorter that the Cu-Cu average  distance, but the number of these 
pairs are 
almost the same. Concerning Ti-Ti pairs, there is a reduction both in its number and distance when 
compared to {\em c}-Cu$_2$Ti.

\begin{table}[p]
\caption{\label{tab1} Structural data determined for {\em a}-Cu$_{64}$Ti$_{36}$. The numbers in 
parenthesis are the errors in the values.}
\begin{tabular}{ccccccccccc}\hline
\multicolumn{11}{c}{EXAFS} \\\hline
& \multicolumn{5}{c}{Cu K-edge} & \multicolumn{5}{c}{Ti K-edge}\\
$R$ factor & \multicolumn{5}{c}{4.2} & \multicolumn{5}{c}{5.4} \\\hline
Bond Type & \multicolumn{2}{c}{Cu-Cu\footnotemark} & \multicolumn{3}{c}{Cu-Ti} & 
\multicolumn{3}{c}{Ti-Cu} & \multicolumn{2}{c}{Ti-Ti}\\
$N$ & 2.5 (0.4)  & 7.0 (1.0) & \multicolumn{3}{c}{4.4 (0.6)}  & \multicolumn{3}{c}{7.8 (1.0)} & 
\multicolumn{2}{c}{2.8 (0.4)} \\
$r$ (\AA) & 2.31 (0.02) & 2.80 (0.01) & \multicolumn{3}{c}{2.42 (0.02)} & \multicolumn{3}{c}{2.42 
(0.02)} & \multicolumn{2}{c}{2.70 (0.01)}\\
$\sigma^2$ (\AA $\times 10^{-2}$) & 0.925 (0.1) & 2.51 (0.2) & \multicolumn{3}{c}{3.47 (0.4)} & 
\multicolumn{3}{c}{3.47 (0.4)} & \multicolumn{2}{c}{1.01 (0.1)}  \\\hline
\multicolumn{11}{c}{{\em c}-Cu$_2$Ti} \\\hline
Bond Type & \multicolumn{2}{c}{Cu-Cu\footnotemark} & \multicolumn{3}{c}{Cu-Ti\footnotemark} & 
\multicolumn{3}{c}{Ti-Cu\footnotemark} & \multicolumn{2}{c}{Ti-Ti\footnotemark}\\
$N$ & 2  & 2 & 2 & 2 & 1 & 4 & 4 & 2 & 2 & 2 \\
$r$ (\AA) & 2.54 & 2.58 & 2.58 & 2.61 & 2.63 & 2.58 & 2.61 & 2.63 & 2.93 & 3.32 \\\hline
\end{tabular}
$^1$There are 9.5 pairs at $\langle r\rangle  = 2.67$ \AA. 

$^2$There are 4 pairs at $\langle r \rangle  = 2.56$ \AA.

$^3$There are 5 pairs at $\langle r \rangle = 2.60$ \AA.

$^4$There are 10 pairs at $\langle r \rangle  = 2.60$ \AA.

$^5$There are 4 pairs at $\langle r\rangle  = 3.08$ \AA.

\end{table}

As it can be seen in Table \ref{tab1}, on the Cu edge, two Cu-Cu subshells and one Cu-Ti shell were 
considered in the first shell in order to find a good fit, whereas on the Ti edge one shell of Ti-Cu and 
one shell of Ti-Ti were needed. We started the fitting procedure using one shell for all pairs. This 
choice did not produce a good quality fit of Fourier-filtered first shells on the Cu edge. Besides that, 
the well-known relations below were not verified in fitting EXAFS data:

\begin{eqnarray*}
c_i N_{ij} &=& c_j N_{ij}\\
r_{ij} &=& r_{ji}\\
\sigma_{ij} &=& \sigma_{ji}
\end{eqnarray*}

\noindent where 
$c_i$ is the concentration of atoms of type $i$, $N_{ij}$ is the number of $j$ atoms located at 
a distance $r_{ij}$ around an $i$ atom and $\sigma_{ij}$ is the half-width of the Gaussian. By 
considering two Cu-Cu sub-shells the quality of the fit on the Cu edge was much improved (see Fig. 
\ref{fig3}). Moreover, the relations above were satisfied.

\begin{figure}
\includegraphics{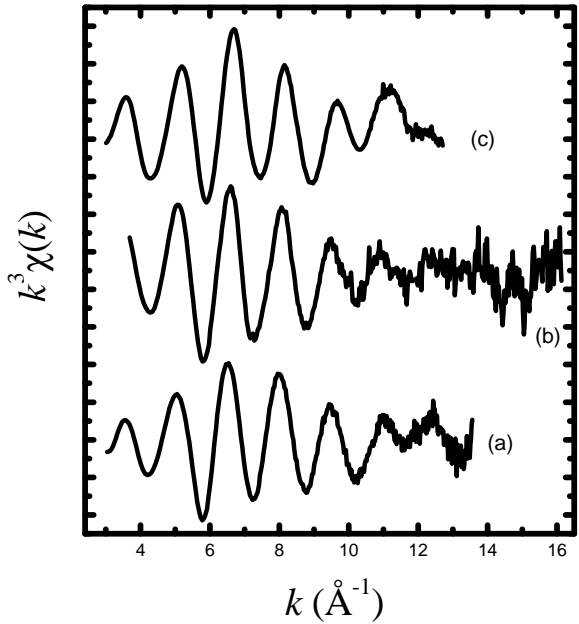}
\caption{\label{fig4} Weighted EXAFS oscillations $k^3\chi(k)$ on the TM K edge for (a) {\em a}-Cu$_{64}$Ti$_{36}$, 
(b) {\em a}-Ni$_{60}$Ti$_{40}$ and (c) {\em a}-Co$_{57}$Ti$_{43}$.}
\end{figure}

It is interesting now to compare EXAFS data found for the three alloys. Fig. \ref{fig4} shows the 
weighted EXAFS 
oscillations $k^3\chi (k)$ obtained on the TM K edges of the three TM-Ti alloys. The oscillations on   
the three K edges are very similar, with small differences in the amplitudes of the 
oscillations.

\begin{figure}[h]
\includegraphics{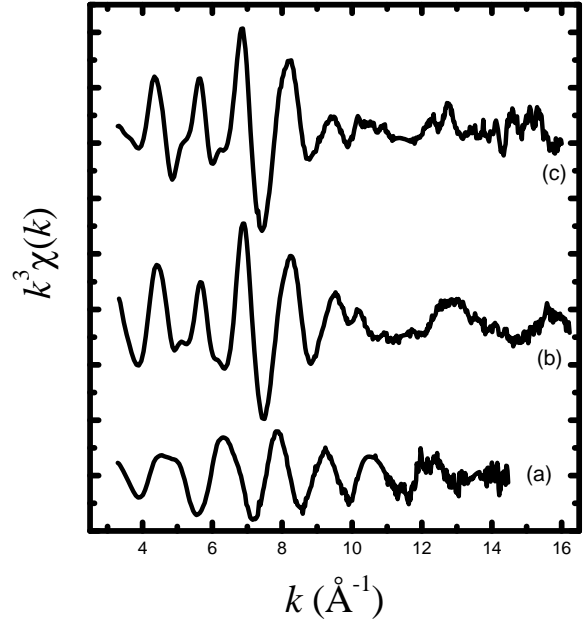}
\caption{\label{fig5} Weighted EXAFS oscillations $k^3\chi(k)$ on the Ti K edge 
for (a) {\em a}-Cu$_{64}$Ti$_{36}$, 
(b) {\em a}-Ni$_{60}$Ti$_{40}$ and (c) {\em a}-Co$_{57}$Ti$_{43}$.}
\end{figure}

Figure \ref{fig5} shows the three weighted EXAFS oscillations $k^3\chi (k)$ on the Ti K edges. 
$k^3\chi_{\rm Ti}(k)$ for {\em a}-Co$_{57}$Ti$_{43}$ and {\em a}-Ni$_{60}$Ti$_{40}$ 
are very similar, indicating that Ti atoms are found in similar chemical environments. However,  
$k^3\chi_{\rm Ti}(k)$ for {\em a}-Cu$_{64}$Ti$_{36}$ shows a different behavior, which reflects the 
differences found in the coordination number determined for Ti atoms in this alloy. For a better 
comparison, structural 
data found for the other alloys are listed in Table \ref{tab2}.

\begin{table}[h]
\caption{\label{tab2} Structural data found for amorphous TM-Ti alloys (average values). The numbers 
in parenthesis are the errors in the values.}
\begin{tabular}{ccccc}\hline
\multicolumn{5}{c}{{\em a}-Co$_{57}$Ti$_{43}$ \cite{KleCoTi}} \\\hline
& \multicolumn{2}{c}{Co K-edge} & \multicolumn{2}{c}{Ti K-edge}\\
$R$ factor & \multicolumn{2}{c}{1.9} & \multicolumn{2}{c}{2.9} \\\hline
Bond Type & Co-Co  & Co-Ti & Ti-Co & Ti-Ti\\
$N$ & 6.0 (0.8) & 6.0 (0.8) & 7.9 (1.0) & 4.9 (0.7)  \\
$r$ (\AA) & 2.50 (0.01) & 2.52 (0.01) & 2.52 (0.01) & 2.98 (0.02)\\
$\sigma^2$ (\AA $\times 10^{-2}$) & 1.45 (0.2) & 4.57 (0.6) & 4.57 (0.6) & 1.36 (0.1) \\\hline\hline
\multicolumn{5}{c}{{\em a}-Ni$_{60}$Ti$_{40}$ \cite{KleNiTi}} \\\hline
& \multicolumn{2}{c}{Ni K-edge} & \multicolumn{2}{c}{Ti K-edge}\\
$R$ factor & \multicolumn{2}{c}{1.5} & \multicolumn{2}{c}{3.5} \\\hline
Bond Type & Ni-Ni  & Ni-Ti & Ti-Ni & Ti-Ti\\
$N$ & 8.8 (1.2)  & 5.2 (0.6) & 7.9 (1.0)  & 5.5 (0.8) \\
$r$ (\AA) & 2.58 (0.02) & 2.55 (0.01) & 2.55 (0.01) & 2.93 (0.02)\\
$\sigma^2$ (\AA $\times 10^{-2}$) & 2.03 (0.3) & 0.33 (0.04) & 0.33 (0.04) & 0.003 (0.0008) \\\hline
\end{tabular}
\end{table}

From Tables \ref{tab1} and \ref{tab2}, it can be seen that the number of TM-TM pairs increases from TM = Co 
to Cu, and the same behavior is seen in the interatomic distances. Concerning Ti-Ti pairs, there is an 
increase in the number of these pairs from Co to Ni, but in {\em a}-Cu$_{64}$Ti$_{36}$ it decreases by 
almost 3 atoms, indicating a different chemical environment around Ti atoms in this alloy. This fact is 
reinforced if Ti-Ti interatomic distances are considered. There is a small reduction in the distance from 
Co to Ni and a very large reduction from Ni to Cu. It should be noted that the average Cu-Cu interatomic 
distance found by EXAFS analysis is larger than that for Cu-Ti pairs. This is an important feature, and it 
is also seen in {\em a}-Ni$_{60}$Ti$_{40}$. In fact, considering TM-Ti pairs, it can be seen that their 
interatomic distances decrease with the TM atomic number. The number of TM-Ti pairs also lowers when TM 
goes from Co to Cu. According to Hausleitner and Hafner \cite{Hausleitner}, for TM-Ti alloys 
this shortening in the TM-Ti interatomic distance is associated with a change of the $d$-band electronic 
density of states in the TM-Ti alloys from a common-band to a split-band form, which depends on the 
difference of $d$-electrons in the TM and Ti atoms. Since this difference increases from Co to Cu, we 
should expect that the shortening effect would be weaker in {\em a}-Co$_{57}$Ti$_{43}$, increasing for 
{\em a}-Ni$_{60}$Ti$_{40}$ and reaching its maximum value in {\em a}-Cu$_{64}$Ti$_{36}$, and this is 
verified in our results. Fukunaga {\em et al}. also found this shortening in 
{\em a}-Ni$_{40}$Ti$_{60}$ produced by MQ \cite{Fukunaga}. It is important to note that this shortening 
effect was also confirmed by RMC simulations \cite{RMC1,RMCA,rmcreview} in {\em a}-Ni$_{60}$Ti$_{40}$ 
\cite{KleNiTi} and {\em a}-Cu$_{64}$Ti$_{36}$ (results to be published elsewhere). This reduction in 
the TM-Ti interatomic distance indicates that the CSRO in TM-Ti alloys increases from Co-Ti to 
Cu-Ti alloys.

\section{Conclusion}

An amorphous Cu$_{64}$Ti$_{36}$ alloy was produced by Mechanical Alloying technique and its local atomic 
structure was determined from EXAFS analysis. We could find coordination numbers and interatomic distances 
for the first neighbors of this alloy. The results obtained were compared to those found in 
{\em c}-Cu$_2$Ti and also with those determined for other two TM-Ti alloys (TM = Co and Ni) produced by 
MA. The most important feature considering these alloys is the decrease in TM-Ti interatomic distance 
as a function of the TM atomic number, as proposed by Hausleitner and Hafner \cite{Hausleitner} to 
TM-Ti alloys. This effect can be associated to the CSRO in the alloys, which increases from Co to Cu.

\begin{acknowledgement}

We thank the Brazilian agencies CNPq, CAPES and FINEP for financial support. We also thank LNLS staff 
for help during measurements (proposal n$^{\rm os}$ XAS 799/01 and XAS 998/01). This study was partially 
supported by LNLS.

\end{acknowledgement}


\end{document}